\documentclass[aps,prl,twocolumn,showpacs]{revtex4}

\input{epsf}

\begin{document}

\title{Quasi-one-dimensional Bose gases with large scattering length}

\author{G. E. Astrakharchik$^{(1,2)}$, D. Blume$^{(3)}$, S. Giorgini$^{(1)}$, and B. E. Granger$^{(4)}$}
\address{$^{(1)}$Dipartimento di Fisica, Universit\`a di Trento and 
BEC-INFM, I-38050 Povo, Italy\\
$^{(2)}$Institute of Spectroscopy, 142190 Troitsk, Moscow region, Russia\\
$^{(3)}$Department of Physics, Washington State University,
Pullman, WA 99164-2814, USA\\
$^{(4)}$Institute for Theoretical Atomic, Molecular and Optical Physics,
Harvard-Smithsonian CFA, Cambridge, MA 02138, USA}
\date{\today}

\begin{abstract}
Bose gases
confined in highly-elongated harmonic traps are investigated
over a wide range of interaction strengths 
using quantum Monte Carlo techniques. 
We find that the properties of a Bose gas under tight transverse 
confinement 
are well reproduced by a 1d model Hamiltonian with 
contact interactions. 
We point out the existence of a unitary regime, 
where the
properties of the quasi-1d Bose gas become independent of 
the actual value of the 3d scattering length
$a_{3d}$. In this unitary regime, the energy of the system
is well described by a hard rod equation of state. 
We investigate the stability of quasi-1d Bose gases
with positive 
and negative $a_{3d}$.
\end{abstract}


\maketitle

In recent years the study of 
quasi-1d
quantum Bose gases has attracted 
a great deal of interest.
Intriguing properties of quasi-1d gases,
such as the exact mapping between interacting bosons and 
non-interacting fermions,  
have been predicted \cite{Girardeau,Olshanii,Petrov}. A 
bosonic gas that
behaves as if it consisted of spinless fermions, a so-called 
Tonks-Girardeau (TG) gas, cannot be described within mean-field theory
since it
exhibits strong correlations; instead, a many-body framework is called for.
While experimental evidence of quasi-1d 
behavior has been reported for bosonic atomic gases under highly-elongated 
harmonic confinement~\cite{EXP1D}, 
TG gases have not been observed yet.
It has been suggested, however, that TG gases
can be realized experimentally 
for either low atomic densities or strong atom-atom interaction 
strengths. 
The 3d
$s$-wave scattering length $a_{3d}$, and hence the 
strength of atom-atom interactions,
can be tuned to essentially any value, including zero and $\pm \infty$,
by utilizing a magnetic atom-atom Feshbach resonance~\cite{EXPFR1,EXPFR2}. 

Utilizing a two-body Feshbach resonance, 
3d degenerate gases with large scattering length $a_{3d}$ have been studied
experimentally and theoretically.
For $a_{3d} \rightarrow \pm \infty$, it is predicted
that the
behavior of the strongly-correlated gas is independent of
$a_{3d}$~\cite{Pethick}.
For {\em{homogeneous}} 3d Bose gases, 
this unitary regime can most likely not be reached
experimentally since three-body recombination is expected to set in when
$a_{3d}$ becomes comparable to the average interparticle distance. 
Three-body recombination leads to cluster formation, and hence
``destroys'' the gas-like state. 
The situation is different for Fermi gases, 
for which the unitary regime has already been achieved 
experimentally \cite{EXPFR2}. 
In this case, the Fermi pressure stabilizes 
the system even for  large $|a_{3d}|$. 
It has been predicted that three-body recombination processes
are suppressed for strongly interacting 1d Bose 
gases ~\cite{gangardt}.
These studies raise the question whether
a {\em{highly-elongated inhomogeneous}} Bose gas, 
that is, an inhomogeneous quasi-1d Bose gas,
is stable as $a_{3d} \rightarrow \pm \infty$.

This Letter investigates the properties of a quasi-1d Bose gas  
at zero temperature over a wide range of interaction strengths 
within a microscopic, highly accurate many-body framework.
We find that the system i) is well described by a 1d model Hamiltonian with
contact interactions and renormalized coupling constant \cite{Olshanii}
for any value of the 3d scattering length
$a_{3d}$; ii) behaves like a TG gas
for a critical positive value of $a_{3d}$; iii)
reaches a unitary regime for large 
values of $|a_{3d}|$,
where the
properties of the quasi-1d Bose gas become independent of 
the actual value of 
$a_{3d}$ and
are similar to those
of a hard-rod gas; and iv) 
becomes unstable against cluster formation
for a critical negative value of $a_{3d}$.                  

\begin{figure}[tbp]
\vspace*{-1.2in}
\centerline{\epsfxsize=3.0in\epsfbox{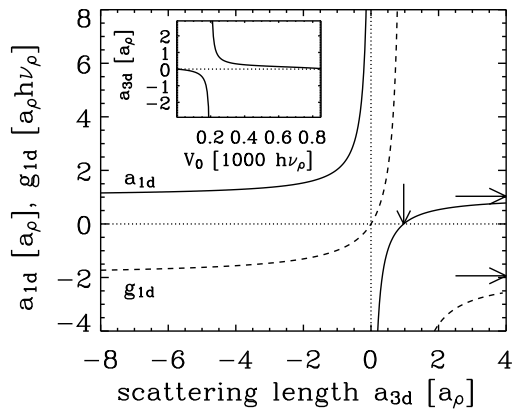}}
\vspace*{-.3in}
\caption{$g_{1d}$ [dashed line, Eq.~(\protect\ref{eq_coupling1d})] 
and $a_{1d}$ [solid line, Eq.~(\protect\ref{eq_scatt1d})] 
as a function of $a_{3d}$.
A vertical arrow indicates the value of
$a_{3d}$ where $g_{1d}$ diverges, $a_{3d}^{c}/a_{\rho}=0.9684$.
Horizontal arrows indicate the asymptotic values of $g_{1d}$
and $a_{1d}$, respectively, as $a_{3d} \rightarrow \pm \infty$
($g_{1d}=-1.9368a_{\rho} \hbar \omega_{\rho}$  
and $a_{1d}=1.0326a_{\rho}$).
Inset: $a_{3d}$ as a function of the well depth $V_0$ for
$V^{SR}(r)$.} \label{fig1}
\end{figure}

Our study is based on the 3d Hamiltonian $H_{3d}$, 
\begin{eqnarray}
H_{3d}= \sum_{i=1}^N \left[ \frac{-\hbar^2}{2m}
\nabla_i^2 + \frac{m}{2} \left( \omega_{\rho}^2 \rho_i^2
+ \omega_z^2 z_i^2 \right) \right] + \sum_{i<j} V(r_{ij}) \;, 
\label{eq_3dtrap}
\end{eqnarray}
which describes $N$ spin-polarized mass $m$ bosons under 
highly-elongated confinement
with $\omega_z=\lambda \omega_{\rho}$, where $\lambda\ll 1$. 
The coordinates $\rho_i=\sqrt{x_i^2+y_i^2}$ and $z_i$ denote, respectively, 
the radial and axial position 
of the $i$-th particle, $r_{ij}=|\vec{r}_i-\vec{r}_j|$ denotes 
the interparticle distance between atom $i$ and $j$, 
and $V(r)$ denotes the two-body interatomic potential. 
We consider two different potentials: i) a purely repulsive 
hard-sphere (HS) potential, $V^{HS}(r)=\infty$ for $r<a_{3d}$, and 
zero otherwise; and ii) a more realistic 
short-range (SR) potential, which can support 
two-body 
bound states, 
$V^{SR}(r)=-V_0/\mbox{cosh}^2(r/r_0)$, with well
depth $V_0$.
For $V^{HS}$, the $s$-wave scattering length $a_{3d}$ 
coincides with the range of the 
potential. For $V^{SR}$, in contrast, 
$r_0$ determines the range of the potential,
while the scattering length $a_{3d}$ is a function of $r_0$ and $V_0$.
In our calculations, $r_0$ is fixed at a value much smaller than the 
transverse oscillator length, $r_0 = 0.1 a_\rho$, where 
$a_\rho=\sqrt{\hbar/m\omega_\rho}$. 
To simulate the behavior of $a_{3d}$ near a field-dependent Feshbach resonance,
we vary the well depth $V_0$ and consequently the
scattering length $a_{3d}$.
Importantly, $a_{3d}$
diverges
for particular values of $V_0$ (the inset of
Fig.~\ref{fig1} shows one such divergence).
At each divergence, a new two-body $s$-wave bound state is pulled
in. 
Our numerical calculations (see below) are performed for
the well depths $V_0$ shown in the inset of Fig.~\ref{fig1}.

We consider situations where the bosonic gas
described by Eq.~(\ref{eq_3dtrap})
is in the 1d 
regime for any value of the 3d scattering length $a_{3d}$, which
implies $N\lambda\ll 1$. 
If the 
range of $V(r)$ is much smaller than $a_\rho$, 
it is predicted~\cite{Olshanii} that the 
properties of the 3d gas are well described by 
the 
1d contact-interaction Hamiltonian $H_{1d}$,
\begin{eqnarray}
H_{1d}=\sum_{i=1}^N \left( \frac{-\hbar^2}{2m}
\frac{\partial^2}{\partial z_i^2} + \frac{m}{2} \omega_z^2 z_i^2 \right) 
+ g_{1d}\sum_{i<j} \delta(z_{ij}) + \nonumber \\ 
 N \hbar\omega_{\rho}  \;,
\label{eq_1dtrap}
\end{eqnarray}
where $g_{1d}$ is an effective coupling constant.
The limit $g_{1d}\to 0$ corresponds to the 
weakly-interacting mean-field regime, while $g_{1d}\to \infty$ 
corresponds to the strongly-interacting TG regime.  
For positive $g_{1d}$ and vanishing confinement
($\omega_\rho=\omega_z=0$), Eq.~(\ref{eq_1dtrap})
reduces to the Lieb-Liniger (LL) Hamiltonian \cite{Lieb},
whose gas-like properties have been studied in detail.
For negative $g_{1d}$ and $\omega_\rho=\omega_z=0$,
Eq.~(\ref{eq_1dtrap}) supports
cluster-like bound states \cite{McGuire}; 
little is known about gas-like states in this case.

Below, 
we solve the many-body
Schr\"odinger equation for the 1d Hamiltonian {\em{with confinement}},
Eq.~(\ref{eq_1dtrap}),
for positive {\em{and}} negative $g_{1d}$, and relate our results
to those for the 3d Hamiltonian, Eq.~(\ref{eq_3dtrap}). 
Considering
two bosons in a highly-elongated geometry that interact 
through a regularized zero-range 
pseudo-potential,
Olshanii \cite{Olshanii} shows that the effective
1d coupling constant $g_{1d}$ 
can be expressed 
in terms of the known 3d scattering length $a_{3d}$,
\begin{eqnarray}
g_{1d}=
\frac{2 \hbar^2 a_{3d}}{m a_\rho^2}\frac{1}{1- A a_{3d}/a_\rho} \,,
\label{eq_coupling1d}
\end{eqnarray}
where $A=|\zeta(1/2)|/\sqrt{2} = 1.0326$.
Alternatively, $g_{1d}$ can be expressed through the effective 
1d scattering length $a_{1d}$,
$g_{1d}= -2\hbar^2/(ma_{1d})$~\cite{Olshanii}, where
\begin{eqnarray}
a_{1d} = - a_{\rho} \left( \frac{a_{\rho}}{a_{3d}} - A \right) \,.
\label{eq_scatt1d}
\end{eqnarray}
For $a_{3d}>0$ (but $a_{3d} \ll a_{\rho}$), $g_{1d}$ approaches the
unrenormalized coupling constant,
$g_{1d}^0=2\hbar^2 a_{3d}/(m a_\rho^2)$, which is
obtained by
averaging the 3d coupling constant $g_{3d}=4\pi\hbar^2a_{3d}/m$ 
over the transverse oscillator ground 
state (see, e.g., Ref. \cite{Petrov}).
For these values of $a_{3d}$, the 1d Hamiltonian given in
Eq.~(\ref{eq_1dtrap}) with $g_{1d}$ replaced by $g_{1d}^0$,
describes a quasi-1d system accurately \cite{DMC}.

For $a_{3d} \gtrsim a_{\rho}$ and for 
negative $a_{3d}$, in contrast, the confinement induced renormalization
becomes important, and the effective 1d coupling constant $g_{1d}$ 
and scattering 
length $a_{1d}$,
Eqs.~(\ref{eq_coupling1d}) and (\ref{eq_scatt1d}), 
have to be used. 
Figure~\ref{fig1} shows $g_{1d}$ (dashed line)
and $a_{1d}$ (solid line) as a function of $a_{3d}$.
At the critical value  
$a_{3d}^c=0.9684 a_\rho$ (indicated by
a 
vertical arrow in Fig.~\ref{fig1}),
$g_{1d}$ diverges
while $a_{1d}$ goes through zero. 
At the 3d resonance, that is for $a_{3d}\to\pm\infty$,
$g_{1d}$ and $a_{1d}$ each 
reach an asymptotic value 
($g_{1d}=-1.9368a_{\rho} \hbar \omega_{\rho}$  
and $a_{1d}=1.0326a_{\rho}$, respectively, indicated by
horizontal arrows in Fig.~\ref{fig1}).
Tuning 
$a_{3d}$ to large values
hence allows a 
unitary quasi-1d regime, where $g_{1d}$ and $a_{1d}$
are independent of $a_{3d}$, to be entered.

We solve the Schr\"odinger equation for the many-body
3d Hamiltonian,
Eq.~(\ref{eq_3dtrap}),
numerically using the diffusion quantum Monte Carlo (DMC)
technique.
Our interest is in the lowest gas-like many-body state.
For $V^{HS}$, the lowest gas-like state of $H_{3d}$ 
coincides with the many-body ground state; to describe this
state we can hence use the ``standard''
DMC technique (see, e.g., 
\cite{Boronat}).
For $V^{SR}$, in contrast, the many-body ground state is a cluster-like bound state.
To describe the lowest-lying gas-like state, i.e., an excited state,
we hence use the fixed-node DMC (FN-DMC) method \cite{Reynolds}. 
For a given many-body nodal surface, 
the FN-DMC method allows the Schr\"{o}dinger equation 
to be solved for approximate excited states. 
For $N=2$, we obtain the exact nodal surface 
by direct diagonalization. 
Assuming that the 
scattering properties between each atom pair are unaltered by 
the presence of other atoms, we then use the $N=2$
nodal surface to
construct
the many-body nodal surface. For dilute quasi-1d gases 
we expect that the FN-DMC approach as implemented here results in highly accurate many-body energies.

\begin{figure}[tbp]
\vspace*{-.2in}
\centerline{\epsfxsize=3.25in\epsfbox{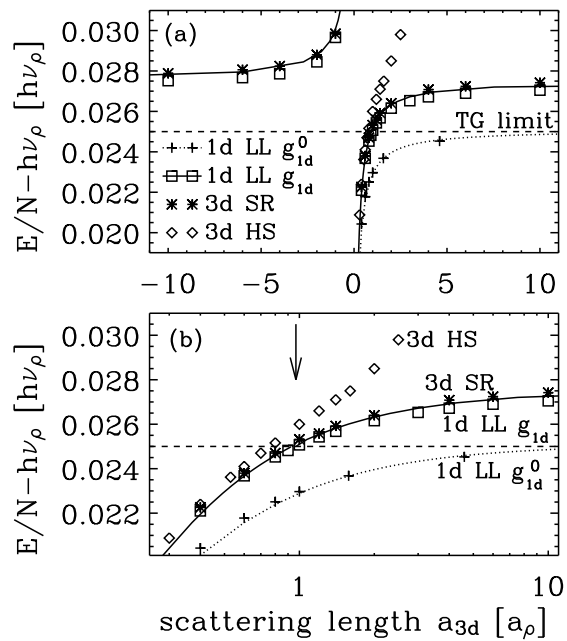}}
\vspace*{-.3in}
\caption{3d DMC energy per particle
calculated using
$V^{HS}$ (diamonds) and $V^{SR}$ (asterisks), respectively,
together with 1d DMC energy per particle calculated using
$g_{1d}$ [squares, Eq.~(\protect\ref{eq_coupling1d})] and 
$g_{1d}^{0}$ (pluses),
respectively, as a function of $a_{3d}$ [(a) linear scale;
(b) logarithmic
scale] for $N=5$ and $\lambda=0.01$.
The statistical uncertainty of the DMC energies is smaller than
the symbol size. 
Dotted and solid lines show the 1d energy per particle calculated
within the LDA for $g_{1d}^0$ (using the LL equation of
state) and for $g_{1d}$, Eq.~(\protect\ref{eq_coupling1d})
(using the LL equation of
state for $g_{1d}>0$, and the HR equation of state for
$g_{1d}<0$), respectively. 
A dashed horizontal line indicates the
TG energy, and a vertical arrow the position where $g_{1d}$, 
Eq.~(\protect\ref{eq_coupling1d}),
diverges.} \label{fig2}
\end{figure}

Figure 2 shows the resulting 3d energy per particle, 
$E/N-\hbar \omega_{\rho}$, as a function of $a_{3d}$ for $N=5$ 
under 
quasi-1d confinement, 
$\lambda=0.01$, for the hard-sphere and the short-range 
two-body potential, $V^{HS}$ (diamonds) 
and $V^{SR}$ (asterisks), respectively. 
For small $a_{3d}/a_{\rho}$, the energies for these 
two two-body potentials agree within the statistical uncertainty. For $a_{3d}\gtrsim a_{\rho}$, however, clear 
discrepancies are visible. The DMC energies for $V^{SR}$ cross the TG energy per particle 
(indicated by a dashed horizontal line), 
$E/N-\hbar\omega_\rho=\hbar\omega_{\rho} \lambda N/2$, very close to the value $a_{3d}^{c}=0.9684 a_{\rho}$ 
[indicated by a vertical arrow in Fig.~\ref{fig2}(b)], 
while the energies for $V^{HS}$ cross the TG energy per 
particle at a notably smaller value of
$a_{3d}$. 

To compare our results obtained for the 3d Hamiltonian, 
$H_{3d}$, with those for the 1d Hamiltonian, $H_{1d}$, 
we also solve the Schr\"odinger equation for $H_{1d}$, Eq.~(\ref{eq_1dtrap}).
For positive values of the coupling constant 
$g_{1d}$ we calculate the many-body ground state energy 
by the exact DMC method. For $g_{1d}<0$, however, the 1d 
Hamiltonian supports many-body bound states,
and, as in the 3d case, we use the FN-DMC method to 
describe the lowest-lying gas-like state.
For $N=2$ and $g_{1d}<0$, the first excited state of the Schr\"odinger equation 
for $H_{1d}$, Eq.~(\ref{eq_1dtrap}), has a node at $z_{12}=a_{1d}$.
To solve the many-body Schr\"odinger equation 
for $H_{1d}$ with negative $g_{1d}$
for the lowest 
gas-like state by the FN-DMC 
technique, we parameterize our many-body nodal surface in terms 
of $a_{1d}$. This many-body nodal surface is expected to be good 
if the density of the gas is low. 

In addition to the 3d energy per particle, Fig.~2 shows 
the resulting 1d energy per particle
obtained by solving 
the Schr\"odinger equation for $H_{1d}$, 
Eq.~(\ref{eq_1dtrap}), for the renormalized 
coupling constant $g_{1d}$ [squares, 
Eq.~(\ref{eq_coupling1d})], and the unrenormalized 
coupling constant $g_{1d}^0$ (pluses), respectively. 
The 1d energies calculated using the two different coupling constants agree well for small $a_{3d}$, while clear discrepancies become apparent for larger $a_{3d}$.
Importantly, the 1d energies calculated using the renormalized 1d coupling constant $g_{1d}$ agree 
well with the 3d energies calculated using  
the short-range 
potential $V^{SR}$ (asterisks) up to very large values of the 3d 
scattering length $a_{3d}$, and also for negative $a_{3d}$. 
In contrast, at large $a_{3d}$ the 1d energies deviate clearly from the 3d energies calculated using  
the hard-sphere
potential $V^{HS}$ (diamonds). 
We conclude that the renormalization of the effective 1d 
coupling constant $g_{1d}$ and the 1d scattering length $a_{1d}$ are crucial to reproduce 
the results of the 3d Hamiltonian $H_{3d}$ when 
$a_{3d}\gtrsim a_\rho$ and when $a_{3d}$ negative. 
Small deviations between the 1d energies calculated using the renormalized 1d coupling constant $g_{1d}$ and the 3d energies calculated using  
the short-range 
potential $V^{SR}$ remain; we 
attribute these to the finite range of $V^{SR}$. 

If the size of the cloud is much larger than the harmonic oscillator length 
$a_z$, where $a_z =\sqrt{\hbar/m \omega_z}$,
it has been shown that the properties of the 1d LL Hamiltonian $H_{1d}$, $g_{1d}>0$, are well described by a simple equation of state using the local density approximation (LDA) \cite{Dunjko}. For $g_{1d}<0$, we instead apply the equation of state for 1d hard-rods (HRs).   Recall that the many-body nodal surface of the lowest-lying gas-like state of $H_{1d}$ with $g_{1d}<0$ is well parameterized by $a_{1d}$. For $z_{ij}>a_{1d}$, the corresponding wave function coincides with that of $N$ 1d hard rods of size $a_{1d}$. For small values of the 1d gas parameter, $n_{1d}a_{1d}\ll 1$, where $n_{1d}$ denotes the linear density, we hence expect that the lowest-lying 
gas-like state of the 1d many-body Hamiltonian with $g_{1d}<0$ is well described by a system of hard rods of size $a_{1d}$.  
The exact energy per particle of the uniform hard rod system is given by $E/N=(\pi^2\hbar^2
n_{1d}^2/6m)/(1-n_{1d}a_{1d})^2$ \cite{Girardeau}. For trapped systems with $N \lambda \ll 1$
we obtain the expansion
\begin{eqnarray}
E/N - \hbar\omega_\rho = \hbar\omega_\rho \frac{N\lambda}{2}\left( 1 + \frac{128\sqrt{2}}{45\pi^2}
\sqrt{N\lambda}\frac{a_{1d}}{a_\rho} + \cdots \right) \;.
\label{eq_HRLDA}
\end{eqnarray}       
The first term corresponds to the energy per particle in the TG regime. 
Similarly, the linear density in the center 
of the cloud is to lowest order given by the TG result, 
$n_{1d}=\sqrt{2N\lambda}/ (\pi a_\rho)$. 
In the unitary 
limit, that is for $a_{1d}/a_\rho=1.0326$, expression (\ref{eq_HRLDA}) 
becomes independent of  
$a_{3d}$, and depends only on $N\lambda$.

Lines in Fig.~\ref{fig2} show the resulting 1d energies per particle
for the LL equation of state ($g_{1d}>0$) as well as the HR equation 
of state ($g_{1d}<0$) calculated within the LDA. Remarkably, the LDA energies  
nearly coincide with the 1d many-body DMC energies 
(pluses and squares, respectively);
finite-size effects play a role only for
$a_{3d}\ll a_\rho$. Our calculations establish for the first time that a simple treatment, i.e., a HR equation of state treated within the LDA, describes trapped quasi-1d gases with negative coupling constant $g_{1d}$ well over a wide range of the 3d scattering length $a_{3d}$. 
For $a_{3d} \rightarrow ^-0$, 
that is, for large $a_{1d}$,
the HR equation of state using the LDA, cannot properly describe trapped
quasi-1d gases, which are expected to become unstable
against formation of cluster-like
many-body bound states for
$a_{1d} \approx 1/n_{1d}$.
We hence investigate the regime with negative $a_{3d}$ in more detail
within a many-body framework.

By comparing with results for the 
3d Hamiltonian, Eq.~(\ref{eq_3dtrap}), 
we have shown above that the 1d Hamiltonian $H_{1d}$,
Eq.~(\ref{eq_1dtrap}), provides
an excellent description of quasi-1d gases. We hence base
our stability analysis
of quasi-1d gases with large effective 1d scattering
length $a_{1d}$ on $H_{1d}$.
We solve the many-body Schr\"odinger equation for $H_{1d}$ 
using the variational quantum Monte Carlo (VMC) method.
Our variational $N$-particle Bijl-Jastrow-type wave function  
consists of one- and two-body terms. The one-body terms
are written as a function of a single variational
parameter $\alpha$, which determines the size of the atomic gas.
The two-body term is parameterized by $a_{1d}$, and 
explicitly accounts for 
correlations. 

\begin{figure}[tbp]
\vspace*{-1.2in}
\centerline{\epsfxsize=3.0in\epsfbox{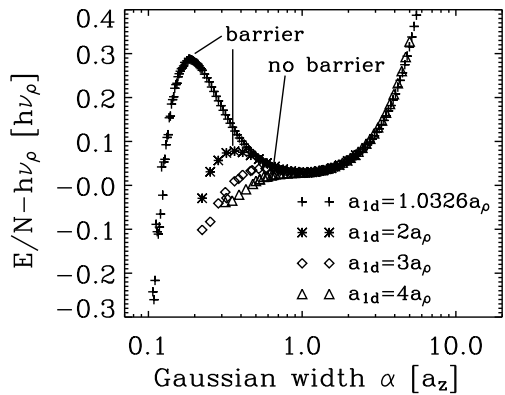}}
\vspace*{-.3in}
\caption{VMC energy per particle as a 
function of the variational parameter $\alpha$ for $N=5$, $\lambda=0.01$, and 
$a_{1d}/a_\rho=1.0326$ (pluses), 2 (asterisks), 3 (diamonds)
and 4 (triangles).}\label{fig3}
\end{figure}

Figure~\ref{fig3} shows the resulting VMC energy per particle
for $N=5$ and $\lambda=0.01$ 
as a function of the variational parameter $\alpha$ 
(Gaussian width) for four
different $a_{1d}$.
For $a_{1d}/a_{\rho}=1$ and $2$, Fig.~\ref{fig3} shows a local minimum 
at $\alpha_{min} \approx a_{z}$. 
The minimum VMC energy
nearly coincides with the
essentially exact DMC energy, which suggests that 
our variational wave function provides a highly accurate 
description of the quasi-1d many-body system.
The energy barrier at $\alpha \approx 0.2 a_{z}$
decreases with increasing $a_{1d}$,
and disappears for $a_{1d}/a_{\rho}\approx 3$.
We interpret this vanishing of the energy barrier as an indication of
instability~\cite{stability}. 
For small $a_{1d}$, the energy barrier separates the 
gas-like state from cluster-like bound states. For larger $a_{1d}$, this
energy barrier disappears and the gas-like state becomes unstable against
cluster formation. 

We additionally performed variational calculations for
larger $N$ and different $\lambda$. We find that the onset
of instability of quasi-1d Bose gases
can be described by the product of the 1d scattering length $a_{1d}$
and the linear density at the trap center $n_{1d}$. 
To be specific, our many-body calculations suggest that
a quasi-1d gas is stable for 
$a_{1d} n_{1d} \lesssim 0.35$, and
becomes unstable for $a_{1d} n_{1d} \gtrsim 0.35$.
Our analysis suggests that the quasi-1d unitary regime can be reached 
experimentally. By tuning the 3d scattering length,
it is further possible to investigate the onset of instability. 
By reducing 
$\lambda$ 
one
should be able to stabilize relatively large
quasi-1d systems. 

In conclusion, we investigated the energetics 
of a Bose gas under highly-elongated harmonic
confinement over a wide range of the 3d scattering length.
We find that the quasi-1d gas can be described by a many-body 
1d model Hamiltonian with 
zero-range interactions and 
renormalized coupling constant. 
For $a_{3d}\to\pm\infty$, the quasi-1d 
gas enters a unitary regime, where all properties of the
system are independent of $a_{3d}$.
In the vicinity of the unitary regime,
the quasi-1d system behaves like a gas of HRs. 
For negative $a_{3d}$, quasi-1d gases become unstable against
cluster formation for a critical value of $a_{1d} n_{1d}$.             

GEA and SG acknowledge support by the Ministero dell'Istruzione,
dell'Universit\`a e della Ricerca (MIUR). DB acknowledges support by
the NSF (grant 0331529)
and 
by the BEC Center at the
University of Trento,
and BEG by the NSF through
a grant to ITAMP.

\end{document}